\documentclass{aa}  

\usepackage{graphicx}
\usepackage{amsmath}
\usepackage{dutchcal}
\usepackage{txfonts}
\usepackage{lipsum}
\usepackage{subcaption}         
\usepackage{lscape}             
\usepackage{placeins}           
   
\newcommand{\msun}{$M_\odot$}

\begin{document}

   \title{Insights from the stability of methanol maser flare properties I: G9.62+0.20E}

   \subtitle{}

%

   \author{S.P. van den Heever\inst{1,2}\thanks{Corresponding author: fanie.vandenheever@nwu.ac.za}
        \and D.J. van der Walt\inst{1}}

   \institute{Centre for Space Research, North West University, Potchefstroom 2531, South Africa 
   \and South Africa Radio Astronomy Observatory (SARAO), Cape Town, South Africa}

   \date{Received September 30, 20XX}

 
  \abstract
   {The brightest 6.7 GHz methanol maser in the Galaxy is associated with the high-mass star-forming region (HMSFR) G009.62+0.20E. Additionally, this methanol maser shows periodic flaring observed at 6.7 and 12.2 GHz. Several models have been hypothesized to explain the periodicity, including periodic accretion, spiral shocks in the gaps of circumbinary disks, pulsating stars, dicke's superradiance, and a colliding wind binary (CWB) system. }
   {We analyzed the time series of the 1.22 km s$^{-1}$ feature of the 6.7 GHz methanol maser spectrum and the 1.25 km s$^{-1}$ feature of the 12.2 GHz methanol maser spectrum over the monitoring duration of roughly 20 years, to determine an updated period, analyze the stability and repeatability of the flares, the temporal behavior of the flare profile, and the duty cycle of the flare profile based on the updated period. From these results, it is argued for or against any of the mechanisms that have been proposed to explain the periodicity.} 
   {Lomb-Scargle (LS) and phase dispersion minimization (PDM) methods were used to determine the methanol maser period at 6.7 and 12.2 GHz. Once the period has been determined, phase-folding or period-folding was used to analyze the flare shape, flare stability, and duty cycle of the flares. }
   {The periods of the 6.7 and 12.2 GHz methanol masers were determined to be $\sim$ 243.0 $\pm$ 0.3 days. The flares are highly repeatable, stable, and show almost identical shapes, as well as their duty cycles. A replicated kernel density estimator (KDE) of the flare shape reproduces the time series very well over a time span of $\sim$ 8000 days. }
   {After careful consideration of the proposed mechanisms for periodic methanol masers, we conclude that all but one, the CWB model, have been ruled out as a possible driver of this flare profile. It has also been established that the CWB model can qualitatively explain the flare profile of the periodic methanol masers. }


   \maketitle
   \nolinenumbers

\section{Introduction}

The star-forming complex G009.62+0.19 has been extensively studied over many years. It consists of several individual objects at different evolutionary stages, as reported by \citet{Garay1993} and \citet{Hofner1996}. The fifth object in this star-forming complex, G009.62+0.20E (hereafter referred to as G962) has a rich history of observations, described as follows: 

\begin{itemize}
    \item \citet{Garay1993} and \citet{Hofner1996} reported that G962 is a hyper/ultra compact ionized region with a size of $\sim$ 0.002 pc. The derived Ultraviolet (UV) luminosity and the ratio between the UV and Far Infrared (FIR) luminosities suggest that G962 harbors a $\sim$ O9.5 type star. 
    \item The continuum spectrum of G962 between 1.5 and 15 GHz has been reported by \citet{Hofner1996} and \citet{Franco2000}; these authors suggest that the source of the continuum is a radially dependent ionized wind or a \ion{H}{II} region with a radially dependent electron density structure. This also has been suggested by \citet{Woode2025}, based on more recent continuum observations that complement the \citet{Franco2000} data. 
    \item \citet{goedhart2003} discovered that the methanol masers at 6.7 GHz and 12.2 GHz associated with G962 show regular/periodic variations in their flux densities with a period of $\sim$ 244.2 days as a result of the long-term monitoring program using the 26m antenna at the Hartebeesthoek Radio Astronomy Observatory (HartRAO). 
    \item \citet{sanna2009} made interferometric observations at 5 different epochs during 2005-2007 with the VLBA to determine the parallax of G962, as well as generate spot maps of the 12.2 GHz masers. These observations complemented the observations of \citet{minier2002a, minier2002b} and \citet{goedhart2005} at 6.7 GHz and 12.2 GHz. From these observations, it was shown that (1) the maser spots are divided into several centralized structures that make up the maser spectrum, (2) no spots are created or destroyed when the masers flare (16 maser spots in all the observations at 12.2 GHz), nor do any structural changes happen and only the flux density changes, and (3) the main periodically variable features of both 6.7- and 12.2 GHz methanol masers seem to be projected at the same position (co-spatial). 
    \item The results of \citet{sanna2009} support the conclusion of \citet{goedhart2005}, which indicates that only the flux of the maser spots changes over the duration of the flare cycle. This strongly indicates that the periodicity is caused by changes in the seed photons or time-dependent pumping of the masers. 
   
\end{itemize}

Since the discovery of the phenomenon of regular/periodic flaring, to date, regular periodic variations in flux density have been discovered from more than 30 sites of high-mass star formation \citep{goedhart2003, Araya2010, szymczak2011, szymczak2015, Szymczak2018, Maswanganye2015, Maswanganye2016, Fujisawa2014}. This set of periodic methanol masers includes a variety of periods and flare profiles. For clarity, the definitions that will be used throughout this work, the period is the time between flare maxima, the flare shape or flare profile describes the temporal behaviour of the flux density over the duration of the period, and the stability is how well the flare profile is described across multiple flares. The first flare profile to be discovered, the profile associated with G962 exhibits an asymmetric shape with a steep rise and a decrease that resemble that of exponential decay, which is also shared by a few more periodic methanol masers. Several hypotheses have been suggested to explain the periodicity and shape of the flare profile. These include periodic accretion in a binary system \citep{Araya2010}, pulsating stars \citep{Inayoshi2013}, rotating spiral shocks in the gaps of the disks of a young binary system \citep{Parfenov2014}, a CWB system \citep{vanderwalt2009, vanderwalt2011, vandenheever2019}, and more recently, Dicke's superradiance \citep[see e.g.][]{Rajabi2023, Rashidi2025, Rashidi2026}. 

Since roughly 2003, the monitoring of G962 with the 26 m Hartebeesthoek radio telescope has accumulated $\sim$ 8000 days (32 cycles) of data. The last update on the period was made in 2014 \citep{Goedhart2014} with data only up until 2008 when the 26 m telescope bearing failed. Thus, a new analysis of the period for the time series of the 6.7 and 12.2 GHz methanol masers could provide a new perspective. A long-term data set like this provides an opportunity to investigate a variety of parameters, such as the stability of the period, the stability of the flare profile, and the duty cycle of the flare profile. The results of such an analysis can provide valuable insight into the most likely mechanism responsible for the periodic behavior. The period is determined using 2 methods, Lomb-Scargle (LS) \citep{Lomb1976,Scargle1982} and phase dispersion minimization (PDM)\citep{Stellingwerf1978}. Based on the determined period, phase folding and flare normalization are applied to analyze the flare stability, the flare shape, the fit to the data over $\sim$ 8000 days, and the duty cycle of the flare profile. 
From the results of the analysis, it is concluded that the colliding-wind binary (CWB) \citep{vanderwalt2009, vanderwalt2011, vandenheever2019} model remains the most likely explanation for the stability of the period and the flare profile of G962.  

\section{Observations}
\label{sec:Obs}
Observations using the 26 \ m Hartebeesthoek Radio Astronomy Observatory (HartRAO) were made with the 4.5 \ cm and 2.5 cm receivers. The receivers are dual, left and right circularly polarized receivers (LCP and RCP), of which the 4.5 \ cm receiver is cryogenically cooled and the 2.5 \ cm receiver is ambient. The data of each receiver and each polarization were calibrated independently relative to Hydra A and 3C123, assuming a flux scale of \citet{Ott1994}. The point-source sensitivity (PSS) per polarization was determined by computing a 30-day average (15 days on either side of the observing date). The beam size of the 4.5 \ cm observation is 7'.0, and the spectral bandwidth used for all observations was 1 MHz, divided into 1024 channels, giving a spectral resolution of $\sim$ 1 kHz per spectral channel, which results in a velocity resolution of 0.044 km s$^{-1}$. For the 2.5 \ cm receiver, the beam size is 3'.7, and a spectral bandwidth of 2 MHz was used, resulting in a spectral resolution of 0.048 km s$^{-1}$.  

G962 is a bright methanol maser, so pointing correction observations are made each time G962 is observed. The uncertainty in the absolute flux is on the order of 5-10\% for all systematic errors that are considered. 
\begin{figure*}[!t]
   \centering
   \includegraphics[width=\textwidth]{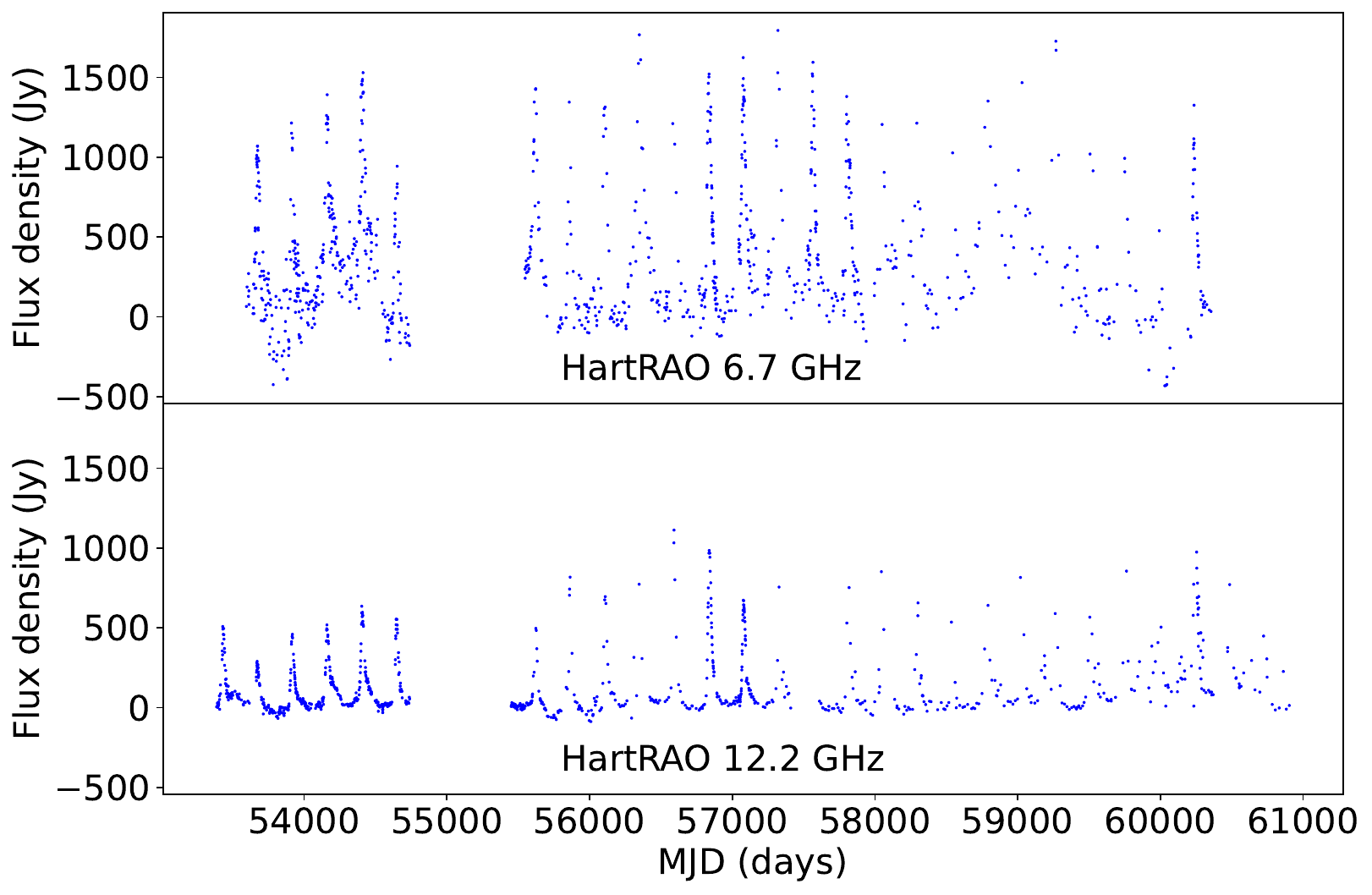}
      \caption{Time series or light curves of the observed methanol masers at the Hartebeesthoek Radio Astronomy Observatory (HartRAO) for G962: Top: the 1.22 km $s^{-1}$ feature of the 6.7 GHz methanol maser spectrum. Bottom: the 1.25 km $s^{-1}$ feature of the 12.2 GHz methanol maser spectrum. }
         \label{fig:ts}
\end{figure*}

\section{Data analysis and results}

The light curves presented in Fig. \ref{fig:ts} were produced by removing long-term variability. This was done by fitting a 5$^{th}$ order polynomial to the data points that represent the quiescent phase between the flares and subtracting the polynomial from the data. The fitting and subtraction of the 5$^{th}$ order polynomial to the data is beyond the scope of this work and will not be discussed further.

\subsection{Period determination}

From these data, G962 shows regular/periodic variations in flux density and \citet{goedhart2003} reported a period of 246.0 $\pm$ 6 days using the LS periodogram analysis. \citet{Goedhart2014} reported a new period, determined from a longer time series (1999-2008), of 244.4 days for the 6.7 GHz methanol masers and 244.0 days for the 12.2 GHz methanol masers. In this work, the updated data for the duration from 2005-2025 were used to re-examine the period determination, using two methods: (1) LS  and (2) PDM. 

\begin{figure}[h!]
   \centering 
   \includegraphics[width=\hsize]{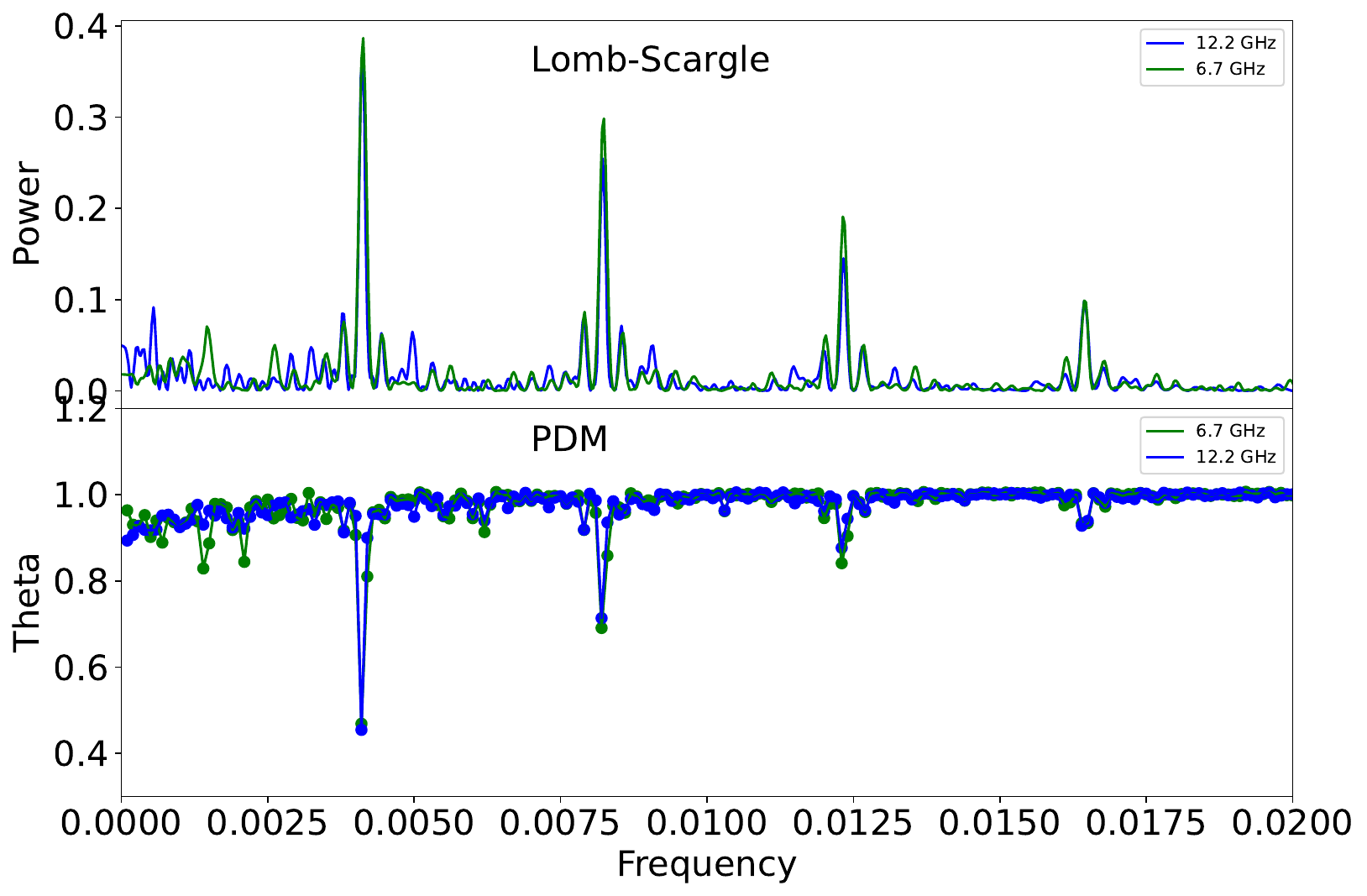}
      \caption{Top: the LS periodogram obtained from the 6.7 GHz (green) and 12.2 GHz (blue) methanol maser light curves. Bottom: the PDM periodogram obtained from the 6.7 GHz (green), and 12.2 GHz (blue) methanol maser light curves. }
         \label{fig:LS-PDM}
   \end{figure}

The LS  periodograms are shown in the top panel of Fig. \ref{fig:LS-PDM} for both the 6.7 GHz (green line) and 12.2 GHz (blue line), showing that the periods obtained from the two methanol masers are essentially equivalent, with values of 243.1 $\pm$ 2.1 days and 243.3 $\pm$ 2.3 days, respectively. Similarly, for the PDM method, as shown in the bottom panel of Fig. \ref{fig:LS-PDM}, the periods of the 6.7 GHz and 12.2 GHz methanol masers are 242.8 $\pm$ 2.3 days and 243.2 $\pm$ 2.2 days, respectively. Thus, the periods and the uncertainty of the periods of these masers are even more constrained \citet[the period originally obtained was 246 $\pm$ 6 days][]{goedhart2003,goedhart2004} as a result of long-term monitoring. In addition, for a sanity check, the time series shown in Fig. \ref{fig:ts} were divided into several sliding windows with an interval of $\sim$ 7 years. The LS and PDM methods were applied to each sliding window to determine its period. \newline

 \begin{figure}[h!]
   \centering
   \includegraphics[width=\hsize]{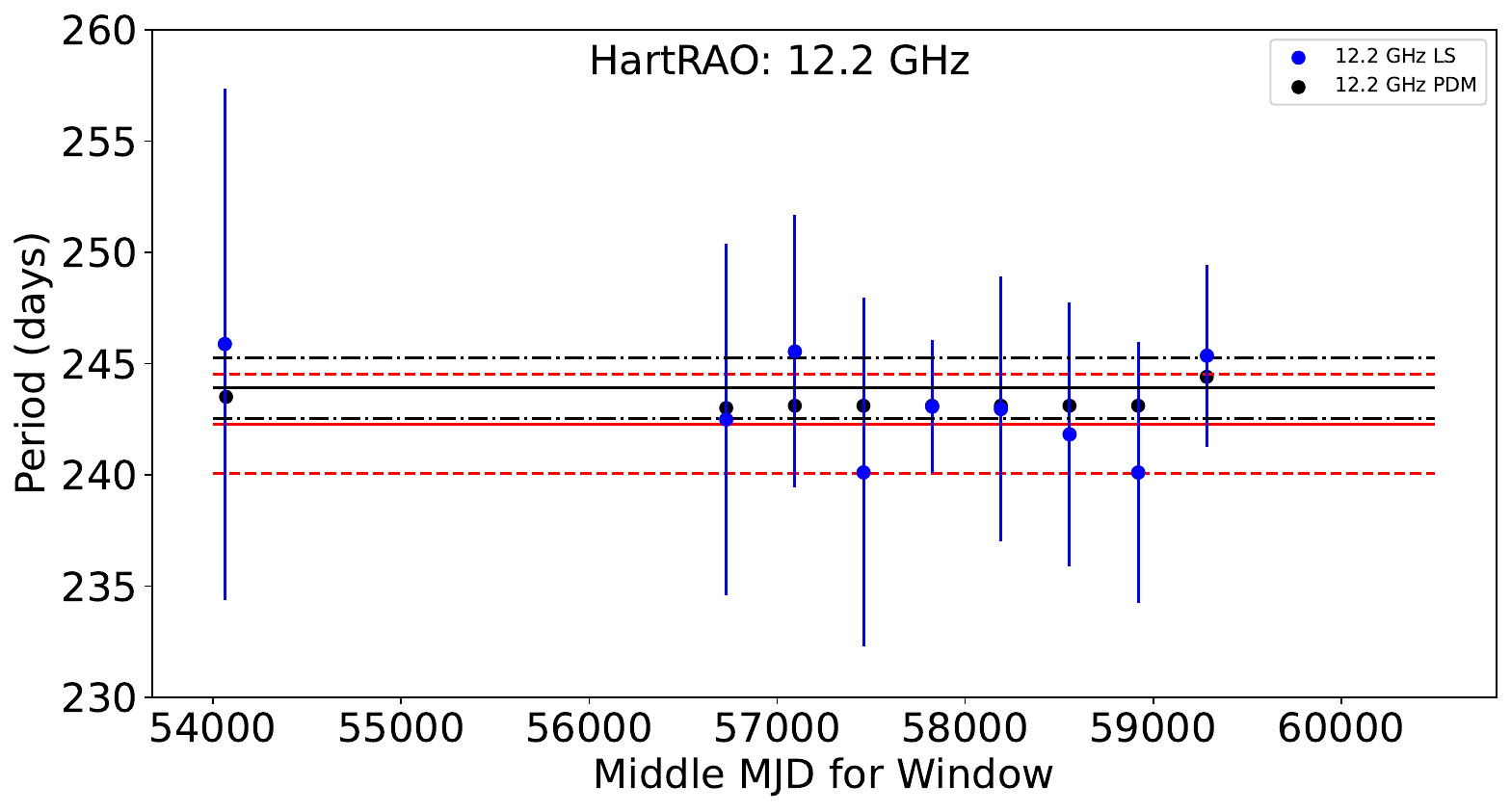}
      \caption{Maser light curve periods obtained as shown in Fig. \ref{fig:LS-PDM}. The horizontal solid and dashed lines represent the period and error margin obtained using the LS method for the entire time series; the same applies to the black solid and dashed lines, but using the PDM method. Additionally, the time series was divided into several sliding windows of roughly 7 years. The blue (LS) and green (PDM) points represent the periods with error margins for each sliding window, where the mean MJD was used to represent the sliding window. It is observed that, within the error margins, both methods constrain the period of G962 very well. }
         \label{fig:LS-PDM-per}
   \end{figure}
The results are shown in Fig. \ref{fig:LS-PDM-per}, where the blue (LS) and green (PDM) points represent the periods of each sliding window at the mean time of each sliding window, respectively. The horizontal lines represent the periods and error bars of the LS (red lines) and PDM (black lines) for the entire time series shown above. This shows that the periods determined from the sliding windows agree well with the period determined for the entire time series. Note that Fig. \ref{fig:LS-PDM-per} only shows the results for the 12.2 GHz time series; however, the results are remarkably similar for the 6.7 GHz time series. Based on these results, the period that will be used throughout the rest of this work is 243.0 days.   

\subsection{Flare stability, shape, and duty cycle}

\subsubsection{Flare stability}

 \begin{figure}[!h]
   \centering
   \includegraphics[width=\hsize]{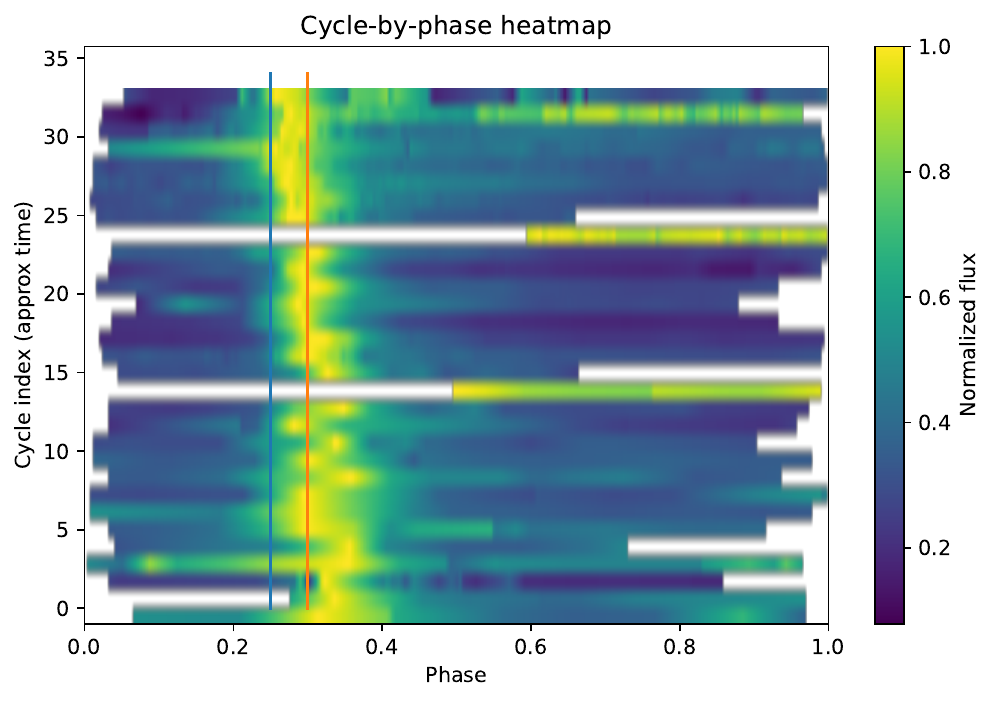}
   \includegraphics[width=\hsize]{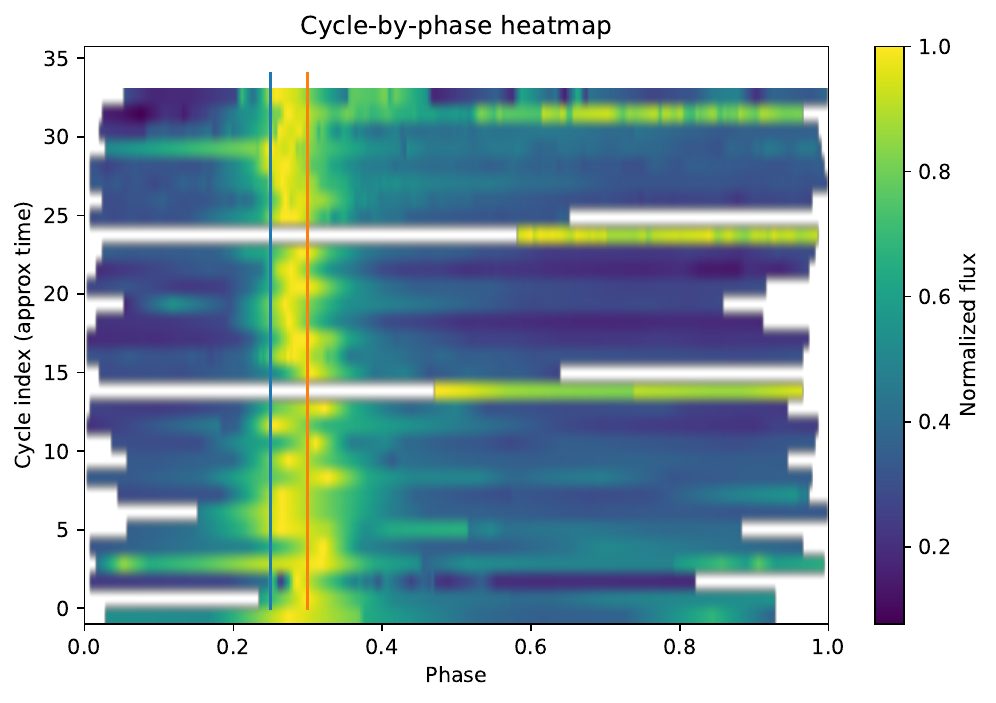}
   \includegraphics[width=\hsize]{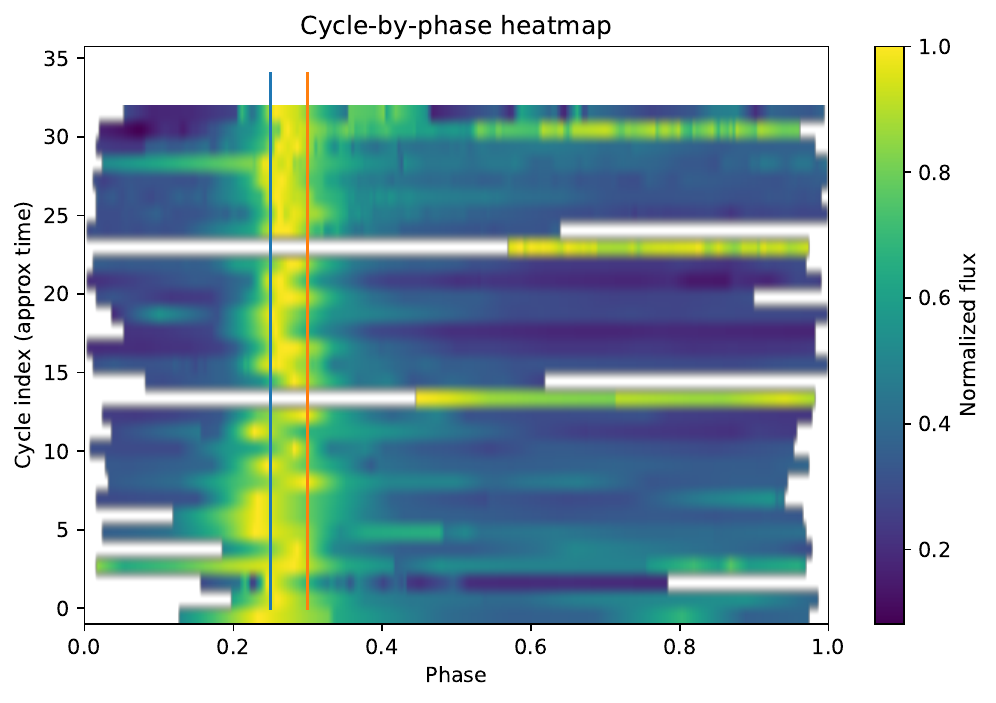}
    
      \caption{Phase-folded flare profiles: The three plots show the phase-folded flare profiles of a period of 243.0 for the 12.2 GHz methanol masers. Middle panel: phase-folded profile for a period of 243.0. Top panel: phase-folded profile for a period of 242.7 days. Bottom panel: phase-folded profile for a period of 243.3 days. These panels show that unless the "true" period is used one will get phase dispersion as the phase drifts. In all three panels the vertical blue and red lines indicate the phase during the flare profile that the flare is at it maximum flux density.}
         \label{fig:PF}
   \end{figure}

With the methanol maser period determined as 243.0 days for both the 6.7 GHz and 12.2 GHz methanol masers, several analyzes can be performed to investigate the repeatability/stability of the flare cycles, their flare shapes, and their flare duty cycles. The first method applied is "period-folding" (also known as "phase-folding"), in which the time series is folded with the determined period, and each observation is converted to a phase during each cycle. The middle panel of Fig. \ref{fig:PF} shows the phase plot as a function of the cycle number for a period of 243.0 days. Note that the reference time of Modified Julian Date (MJD) 52880.0 (August 29, 2003) was used to produce the phase plot, and this reference time will be used throughout this work. Fig. \ref{fig:PF} shows remarkable regularity in the phase and duration of the flares. The vertical red and blue lines represent an interval of $\sim$ 0.04 in phase (equivalent to $\sim$ 9.72 days, which is roughly the time duration of the flare maximum). To test the stability of the flare period, we slightly perturbed the period and set it to 242.7 days (top panel) and 243.3 days (bottom panel). The effects of changing the period are clearly visible in the phase plots, where the flare peaks are dispersed over the flare duration as a result of the "incorrect" period. This shows that although the period uncertainty is on the order of 2 days, the phase plots strongly suggest that the period uncertainty is $\leq$ 0.3 days. 

\subsubsection{Flare shape and duty cycle}
 \label{sec:results}

\begin{figure}[!h]
   \centering
   \includegraphics[width=\hsize]{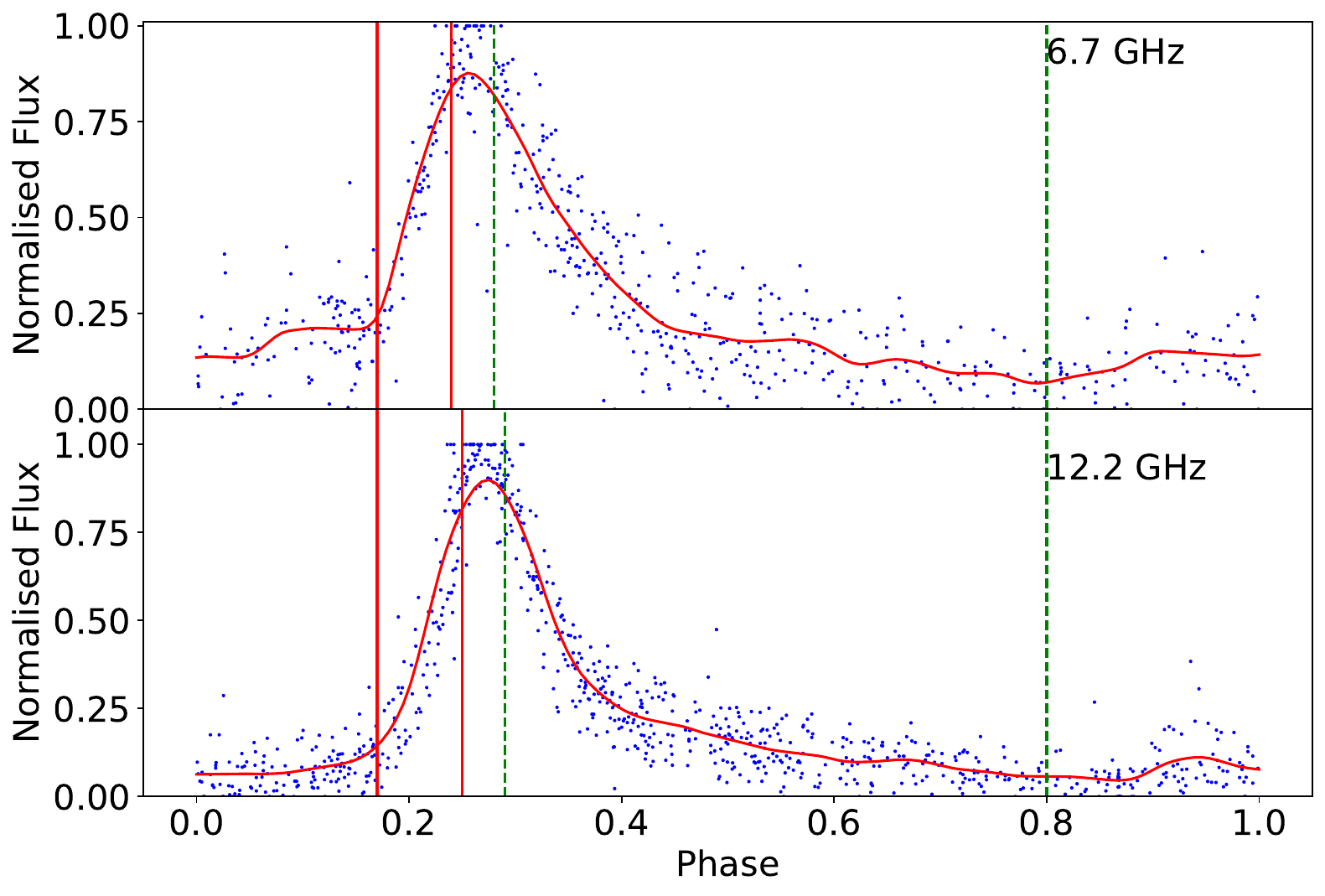}
      \caption{Top panel: 6.7 GHz methanol maser of G962. Bottom panel: 12.2 GHz methanol maser of G962. As seen, both the 6.7 GHz and 12.2 GHz flare profiles show remarkable similarity, and in addition, all the flares of both 6.7 GHz and 12.2 GHz also show remarkable similarity. In both panels the red vertical lines show the start and the end of the raise of the flare. There is a slight difference between the 6.7 - and 12.2 GHz, however, the difference can simply be due to a selection bias. Between the second red line and the first dashed green line, is seen as the time during flare maximum, and the time between the dashed green lines is the time for which it takes the flare to decay to the quiescent state. }  

         \label{fig:PF-KDE}
   \end{figure}

To investigate the repeatability of the period and the shape of each  of the flare profiles of G962, every flare cycle is normalized using:
\begin{equation}
    F_{i,N} = (\frac{F_{i} - F_{min}}{F_{max} - F_{min}})
    \label{eq:1}
\end{equation}

which normalizes each flare between 0 and 1. F$_{i}$ is the i$^{th}$ data point, F$_{max}$ and F$_{min}$ are the data points with the maximum and minimum flux densities for that cycle, and F$_{i,N}$ is the normalized flux density of the $i^{th}$ data point, respectively. As mentioned above, time t$_{0}=  MJD \ 52880$ is the reference time used throughout the work, and the MJD at phase = 0 is simply an integer multiple of the period plus the reference MJD. The flares are folded to fit between 0 and 1, and if the period is correct, the peaks of the flares should all be located at the same phase, as seen in the middle panel of Fig. \ref{fig:PF}. Thus, to clarify, the height of the flare profile is normalized between 0 and 1, and the observation epoch is also normalized between 0 and 1. In addition, note that, as a result of this approach, if no sampling was done close to the peak of the flare, this plot would have outliers; because the flares would be normalized to 1 far from the expected flare maximum. Fig. \ref{fig:PF-KDE} shows the folded normalized flare profiles of both the 6.7 GHz and the 12.2 GHz profiles. The flares with no sampling close to the flare maximum were removed from this analysis as they produce outliers as a consequence of the normalization used. Although the distribution of data points varies for each flare cycle and phase, the flare shape exhibits remarkable repeatability. The ensemble of these points in each flare cycle can be represented as an ''average'' flare profile using the tri-weight kernel density estimator (KDE) 
\begin{equation}
\hat{f}_h(x) = \frac{1}{nh}\sum_{i=1}^nK\left(\frac{x - x_i}{h}\right)
\end{equation}
with
\begin{equation}
    K(u) = \frac{35}{32}(1 - u^2)^3,~~~\vert u \vert \le 1
\end{equation}
\citep[see e.g.][]{Nadaraya1964,Watson1964}, where $\hat{f}_{h}$ is the mean density function for the data points based on the defined kernel density function $K$, with bandwidth (h) and range $\lvert u \rvert = 1$. This ``average'' flare profile is represented by the red lines in the top panel (6.7 GHz) and the bottom panel (12.2 GHz) of Fig. \ref{fig:PF-KDE}. The KDE estimation seems to describe the flare profiles very well. 

The stability of the flare profile can also be investigated through repeatability of its duty cycle, where the duty cycle has three intervals; (1) the rise time and the time during flare maximum, (2) the decay time from flare maximum to the quiescent level, and (3) the time in the quiescent phase before the next flare starts. These three intervals of the flares are indicated by vertical solid red and green dashed lines in Fig. \ref{fig:PF-KDE}, except for the quiescent phase, which simply connects the green dashed line to the red line as the flare restarts. From left to right, the vertical lines represent a phase=0.18 (red), phase = 0.26 (red), phase = 0.30 (green), phase = 0.8 (green). phase = 0.18 - 0.30 represents (1) the rise time plus the time during flare maximum, (2) phase = 0.3 -0.8 represents the decay of the flare, and 0.8 - 0.18 represents the quiescent phase of the flare. The time during flare maximum is indicated by the interval between the second solid red line (phase=0.26) and the first dashed green line (phase = 0.30). The decay of the flares is indicated by the time between the two dashed green lines, and as mentioned, the times during the quiescent phase are described by the interval between the second green dashed line and the first solid red line on the left. As seen, the rise time differs slightly for the 6.7 GHz (top) and 12.2 GHz (bottom) cases. The duration is roughly 7\% (8\%) of the period. This equates to 17.01 days for the 6.7 GHz maser and 19.44 days for the 12.2 GHz maser, respectively. The time spent at flare maximum is equal for both masers and extends $\sim$ 4\% of the period, which is 9.72 days. However, for phase (1) the total time is roughly 17-19 days + 9.72 days, which equals 27 days for 6.7 GHz and 29 days for 12.2 GHz, respectively. The decay part of the flares, on the other hand, extends for half the period ($\sim$ 50\% of the period; phases 0.3-0.8, which is 121.5 days. Thus, this shows that the decay of the flare for both the 6.7 GHz and 12.2 GHz methanol maser flares exhibits remarkable similarity and repeatability over a 20-year ($\sim$ 30 cycles) time span. 

\subsection{Light curve comparison}

\begin{figure*}[h]
   \centering
   \includegraphics[width=12cm]{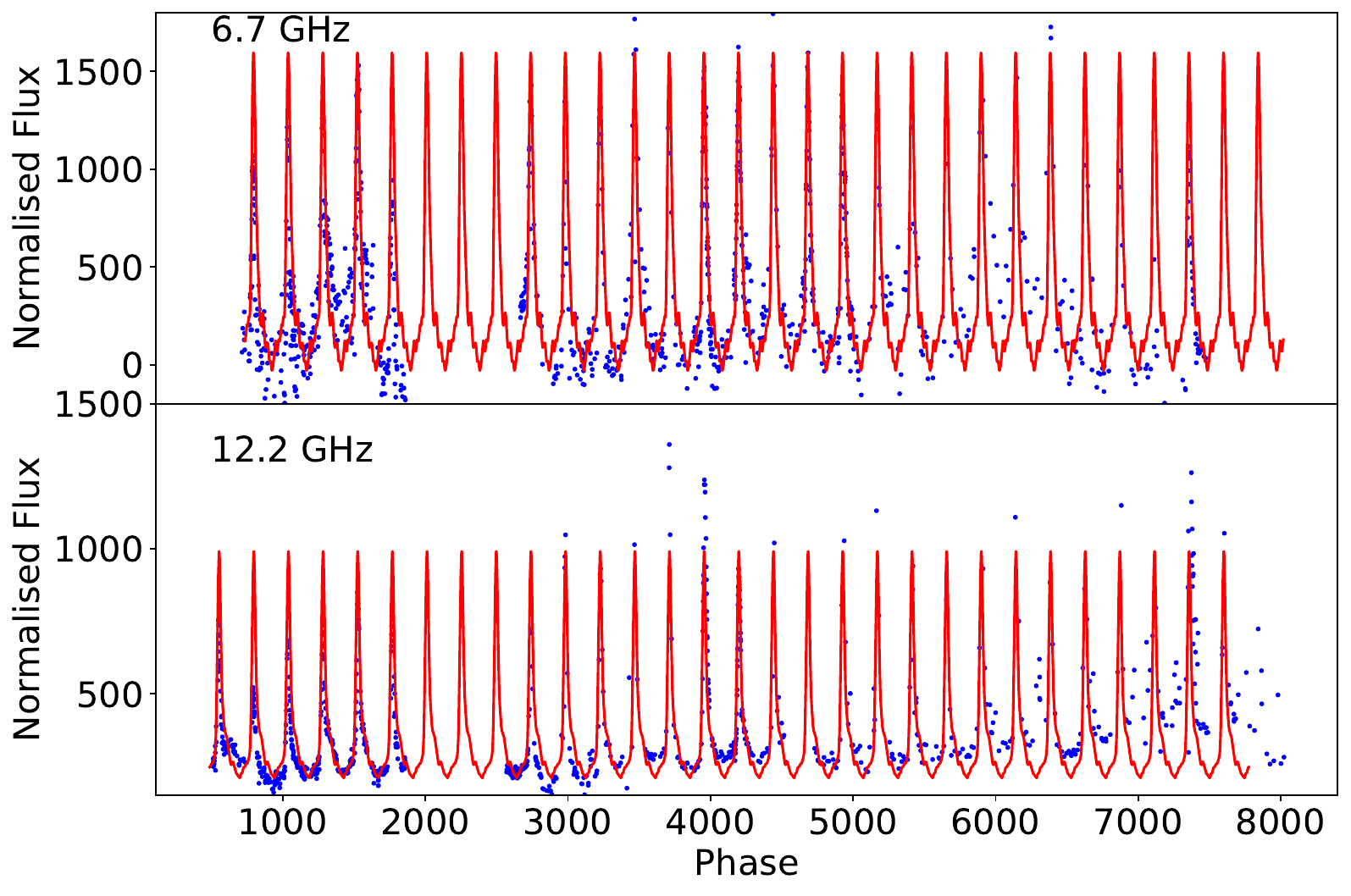}
      \caption{Top panel: The KDE fit of the 6.7 GHz flare profile scaled back to fit to the 6.7 GHz time series. The KDE fit is multiplied by a factor of 3200 and then 700 is subtracted. Bottom panel: The KDE fit of the 12.2 GHz flare profile scaled back to fit the 12.2 GHZ time series. The KDE fit is multiplied by a factor of 2000, and then 100 is added. }
         \label{fig:KDE-fit}
   \end{figure*}

In order to compare the average profile determined using KDE, the KDE profile is duplicated and adjusted to fit to the first observed flare maximum. The average profile is also scaled back to flux density to fit the original data; in this case, the normalized profile is multiplied by 3200 minus 700 for 6.7 GHz and 2000 plus 100 for 12.2 GHz, respectively. As seen in Fig. \ref{fig:KDE-fit}, the average profile fits the data remarkably well. These results show that the flare profiles are highly repeatable over a long time. 
  
\section{Discussion}

The results clearly show that the period and flare profiles of both the 6.7- and 12.2-GHz masers in G962 remain very stable over more than 30 cycles. Such observed stability, flare shape, and duty cycle are characteristics that must be addressed in any proposed explanation of the mechanism underlying the periodicity and flare shape. We present a discussion of the implications of the results presented above on the two possible mechanisms: (1) periodicity as a result of maser pumping, which includes two mechanisms (I) FIR pumping and (II) Dicke's superradiance, and (2) the CWB model. 

Class II methanol masers have been well established to be radiatively pumped by mid-infrared (MIR) and far-infrared (FIR) photons \citep{Sobolev1997} emitted by warm dust. The primary source of dust heating in high-mass star-forming regions is non-ionizing radiation from the central hot star, which is reprocessed to longer wavelengths through repeated absorption and re-emission by dust in the circumstellar envelope or disk. 

\subsection{Comments on possible changes in FIR pumping}
\label{section:FIR}
The proposed mechanisms for changes in the FIR pumping rate are, for example, stellar pulsations \citep{Inayoshi2013,sanna2015, Tanabe2024, Harajiri2026}, periodic/modulated accretion events \citep{Araya2010, szymczak2011, Olech2022} and rotating spiral shocks \citep{Parfenov2014}. \citet{vanderwalt2016} already argued at length that a pulsating star cannot be the underlying mechanism for periodic masers in G962, since a pulsating star has no quiescent phase, as is the case for the periodic methanol masers in G962. Apart from the arguments in \citet{vanderwalt2016}, we also want to point out that there are additional reasons why stellar pulsation is an unlikely driving mechanism for the periodic masers in G962. 

Starting with equation 24.18 of \citet{Draine2011}, the steady-state temperature of a dust grain at a distance $d$ from a star depends on the luminosity, L$_{\ast}$, of the star as 
\begin{equation}
T_d(d) \propto L_\ast^{1/(4 + \beta)} d^{-2/(4+\beta)}
\label{eq:tdust}
\end{equation}
where $\beta$ is the power-law index for the frequency dependence of the dust absorption cross section, $Q_{\mathrm{abs}}(\nu) = Q_0(\nu/\nu_0)^\beta$ \citep{Tielens2005}, with $1 \leq \beta \leq 2$. A change, $\Delta L_\ast$, in the stellar luminosity produces a change, $\Delta T_d$, in dust temperature such that 
\begin{equation}
\frac{\Delta T_d}{T_d} = \frac{1}{4 + \beta}\frac{\Delta L_\ast}{L_\ast}
\label{eq:deltddell}
\end{equation}
Thus, a 10\% increase in dust temperature requires a 50-60\% increase in luminosity. \citet{Hofner1996} classified the central star of G962 as an O9.5 type star with an associated hyper/ultra-compact \ion{H}{ii} region. It is first noted that the bloated-star model of \citet{Inayoshi2013}, which implies a stellar surface temperature of a few thousand K, therefore does not apply to G962. An increase in luminosity must then be due to an increase in the star's effective temperature, not its radius. A 50\% increase in stellar luminosity implies that the central star's radiative properties should oscillate, with a period of 243 days, between those of an O9.5 and an O8.5-type star \citep{Sternberg2003} with an accompanying change by a factor of 2.24 in $Q_H$ (ionizing photons $s^{-1}$). Since the Str\"omgen radius depends on $Q_H^{1/3}$, the required change in luminosity results in a $\approx$ 30\% change in the Str\"omgen radius. Such regular changes in the size of the \ion{H}{II} region would certainly have been noted observationally. Since no such changes have been reported, it is concluded that stellar pulsation as a mechanism to explain the periodic masers in G962 can be ruled out. 

Although not as well established as for the flare profile, it is necessary to note that, in general, the periods of pulsating stars are not stable \citep[see e.g.][]{Templeton2005, Neilson2016}. The results presented above show that the periods of the 6.7- and 12.2-GHz masers in G962 are very stable. However, it is unclear to what extent the pulsation period of the bloated star in the model of \citet{Inayoshi2013} remains stable over 8000 days.   

\subsection{Comments on superradiance modelling of periodic maser flares} Several papers recently appeared to explain maser flaring events, as well as some periodic masers, with Dicke's superradiance theory \citep[see e.g.][]{Rajabi2019, Rajabi2023, Rashidi2025, Rashidi2026}. As far as periodic masers are concerned, the periodic methanol maser flare profiles at 6.7 GHz and 12.2 GHz for G962 have been remarkably well reproduced by \citet{Rajabi2023} and for the 6.7 GHz masers in G22.356+0.066 by \citet{Rashidi2025} within the framework of maser superradiance. Despite this apparent success in reproducing the flare profiles, there are several concerns about the physical applicability of the model to G962. 

\subsubsection{Stellar pulsation origin of the pump pulse }
A central aspect of the proposed superradiance origin of the periodic masers is the source of the pump pulse. In the case of G962, \citet{Rajabi2023} use a pulse duration of $3.5 \times 10^5$s $\approx$ four days. For a period of 243 days, this means that the pulse has a duty cycle of $\sim$ 0.017. \citet{Rajabi2023} states that the superradiance transient response of the system carries no information about the nature of the pump source or its periodicity. In this regard, as these authors state, the periodicity of the pump signal may be due to stellar pulsations from accretion episodes, as suggested by \citep{Inayoshi2013}. It was argued extensively by \citet{vanderwalt2016}, based on the duration of the quiescent phase between flares, that the periodic behaviour of the masers in G962 cannot be explained by a pulsating star. The situation is much worse in the superradiance scenario proposed by \citet{Rajabi2023}, where the quiescent phase of the pumping pulse lasts 239 days. Also, as already indicated in Section\,\ref{section:FIR}, the exciting star of G962 is an ionizing star and therefore does not have the properties of a bloated star as discussed by \citet{Inayoshi2013}, which further rules out stellar pulsation as the source of the pump pulse.  

\subsubsection{Dust heating by emission from the shocked gas in a CWB}
\label{secc:dust}
\citet{Rajabi2023} also considers a colliding-wind binary system \citep{vanderwalt2011,vanderwalt2016} as a possible source of the pump pulse. These authors suggested that ionizing radiation from the shocked gas could be a source of changes in the local dust temperature and, subsequently, in the pump pulse. There are three factors to consider when evaluating a CWB as the underlying mechanism for the pump pulse in the superradiance theory to explain the periodic masers in G962, viz. (1) what orbital parameters are required to produce a pulse with a duration of four days, (2) can the emission from the shocked gas produce a significant increase in dust temperature, and (3) can the FIR/MIR emission from the shocked gas act as a pump pulse.

To illustrate the above, consider a binary system in which, at periastron, the separation between the two stars is $r_p$. From the conservation of angular momentum, it follows that at periastron $r_p^2\frac{d\theta}{dt} = \sqrt{GMa(1-\epsilon^2)}$, where $ M$ is the total mass of the system, $a$ the semi-major axis and $\epsilon$ the eccentricity. Using $r_p = a(1-\epsilon)$ and $\sqrt{GM/a^3} = \frac{2\pi}{P}$, with $P$ as the period, it follows that

\begin{equation}
    \frac{d\theta}{dt} = \frac{2\pi}{P}\frac{\sqrt{1 + \epsilon}}{(1-\epsilon)^{3/2}}
\end{equation}

If we define the time scale, $\mathbf{t_{peri}}$, for the change in $\theta$, and therefore in the separation between the two stars, at periastron as 
\begin{equation}
    t_{peri} = (\frac{d\theta}{dt})^{-1}
    \label{eq:taudOmega}
\end{equation}    
then
    \begin{equation}
    t_{peri} = \frac{P}{2\pi}\frac{(1 - \epsilon)^{3/2}}{\sqrt{1+\epsilon}}
    \label{eq:tau}
    \end{equation}
With $t_{peri}$ = 4 days (the length of the pulse suggested by \citet{Rajabi2023}) and $P$ = 243.0 days, one finds $\epsilon = 0.735$. Taking into account the results of, for example, \citealt{Kobulnicky2014} and \citealt{Villasenor2021}, high-mass binary systems with $\epsilon > 0.6$ are rare. 

Assume further that the binary system, with an eccentricity of 0.735, consists of an O9- and a B0 star. The total mass of the system is $\approx$ 35\msun{}. For such a system, the semi-major axis, $a$, equals 2.5 AU and the periastron separation is 0.66 AU. For the O9 star we assume a wind mass loss rate  $\dot{M}_\mathrm{O9} = 10^{-7}$\msun{}$\,\mathrm{yr^{-1}}$ and a wind speed $\varv_\mathrm{O9} =$ 2000 km$\,\mathrm{s^{-1}}$. For the B0 star the corresponding values are, respectively, $\dot{M}_\mathrm{B0} = 10^{-8}$\msun{}$\,\mathrm{yr^{-1}}$ and $\varv_\mathrm{B0} = $1500 km$\,\mathrm{s^{-1}}$. Following \citet{Stevens1992}, we derive the following results. In the case of a strong adiabatic shock, we have
\begin{equation}
    T_{\mathrm{shock}} = \frac{3}{16}\frac{\mu m_p}{k}\varv^2
\end{equation}
Then, for the O9 star wind shock, $T_{\mathrm{O9}}\approx 5.6 \times 10^7$K and $T_{\mathrm{B0}}\approx 3.2 \times 10^7$K for the wind shock of the B0 star. At these temperatures, emission from the shocked gas is dominated by X-rays, with some ultraviolet and extreme ultraviolet emission, and negligible FIR/MIR emission (as shown below). X-ray interaction with a dust grain is through either Compton scattering by an electron or by ejecting an electron from a deep atomic shell or photo-ionization of one of the constituent atoms \citep{Kruegel2008}. Although the energies of individual X-ray photons are significantly higher than those of non-ionizing lower energy photons, the luminosity of lower energy photons (which directly and through reprocessing heats the dust) produced by the high-mass star is much higher than that of the X-rays from the hot shocked gas \citep[see Fig. 2 of][]{vandenheever2019}. Since the temperature of a dust grain depends on the total luminosity of the radiation field to which it is exposed, it follows that  X-rays do not contribute significantly to dust heating. Ionizing photons are strongly absorbed in the hyper-compact \ion{H}{II} region and are thus also ineffective in heating the dust. Direct heating of the dust through emission from the shocked gas can therefore be ruled out. 

A more realistic scenario is one in which the shocked gas cools radiatively at periastron passage. In this case, \citet{vandenheever2019} showed that the emission of lower-energy ionizing photons increases at periastron passage, not the X-ray emission. However, it is the lower-energy ionizing photons that are absorbed, giving rise to small changes in the position of the ionization front rather than escaping from the \ion{H}{II} region to heat the dust.   

The FIR/MIR emission from the shocked gas is also too faint to directly affect maser pumping. Defining the wind momentum ratio as 
\begin{equation}
    \eta = \frac{\dot{M}_{\mathrm{B0}}\varv_{\mathrm{B0}}}{\dot{M}_{\mathrm{O9}}\varv_{\mathrm{O9}}},
\end{equation}
it follows that $\eta \approx 0.075$. The shock distance from the O9 star is then

\begin{equation}
    r_{\mathrm{O9}} = \frac{r_p}{1 + \sqrt{\eta}} \approx 0.78r_p. 
    \label{eq:bal}
\end{equation}
The wind of the O9 star, therefore, dominates the density and determines the emission measure required to calculate optical depths. To calculate the number density at the wind shock of the O9 star, we use 
\begin{equation}
    \rho = \frac{\dot{M}}{4\pi r^2 \varv}
    \label{eq:rho}
\end{equation}
With $r = r_{\mathrm{O9}}$, the number density at the O9 star wind shock is $\approx 4 \times 10^7~\mathrm{cm^{-3}}$. Behind the shock, the number density increases fourfold, so that within the shocked region, the number density is $\approx 1.6 \times 10^8~\mathrm{cm^{-3}}$. Assuming the thickness of the shock is $~0.1r_p $, the emission measure through the center of the shocked region is $\mathrm{EM} = 6.4\times 10^9~\mathrm{cm^{-6}\,pc}$. The optical depth for a thermal hydrogen plasma is given by 
\begin{equation}
    \tau_\nu = 8.235\times10^{-2}\mathrm{T_e}^{-1.35}\nu^{-2.1}EM
\end{equation}
with $\nu$ in GHz \citep{Rohlfs2000}. For $\nu$ = 100 GHz, $\tau = 1.4\times 10^{-6}$, which means that the shocked gas is highly optically thin at 100 GHz. The brightness temperature of the free-free continuum at 100 GHz is therefore $\approx$ 67 K and rapidly decreases for higher frequencies due to the $\nu^{-2.1}$ dependence of the optical depth. For example, at 1000 GHz, the free-free brightness temperature is only 0.5K and is significantly lower in the far-infrared regime. 

Therefore, it is highly unlikely that the free-free continuum from the hot shocked gas in a colliding-wind binary system can be the source of photons for the short periodic pumping pulses required in the superradiance scenario to explain the periodic masers in G9.62+0.20E.

\subsubsection{Interpretation of rate equation for the inversion population density}
An important point made by \citet{Rajabi2023} is that the superradiance transient response provides no information about the nature of the astronomical source (a pulsating star or a CWB) responsible for the pump and periodic maser flaring. Why this is so and why it poses a problem for interpreting the periodic flaring in G962 are shown in the following simplified analysis.   

Equation 7 of \citet{Rajabi2023} is the rate equation for half of the inversion population density, i.e. for $(n_u-n_l)/2$, where $u$ and $l$, respectively denote the upper and lower levels of the maser transition. The rate equation for $\Delta n = n_u-n_l$ can be derived from the standard rate equations. Consider, therefore, the standard rate equations for the upper and lower levels of the maser transition, viz. 
\begin{equation}
    \frac{dn_u}{dt} = -A_{ul}n_u -U_{ul}(t)B_{ul}n_u + U_{ul}(t)B_{lu}n_l -\Gamma_u(t) n_u + \Lambda_u(t)
    \label{eq:rate1}
\end{equation}
\begin{equation}
    \frac{dn_l}{dt} = A_{ul}n_u + U_{ul}(t)B_{ul}n_u - U_{ul}(t)B_{lu}n_l -\Gamma_l(t) n_l + \Lambda_l(t)    
    \label{eq:rate2}
\end{equation}
where $A_{ul}$ is the Einstein A-coefficient for spontaneous emission $u \rightarrow l$, $B_{ul}$ and $B_{lu}$ are the associated Einstein B-coefficients respectively for stimulated emission and absorption, $U_{ul}$ the energy density in the external radiation field at frequency $\nu = E_{ul}/h$, $\Gamma_{l,u}(t)$ the time dependent total loss rate out of levels $l$ and $u$ to all other levels, and $\Lambda_{l,u}(t)$ the corresponding time dependent pump rates into levels $l$ and $u$ from all other levels. Equations \ref{eq:rate1} and \ref{eq:rate2} constitute a three-level system where the last two terms on the right-hand side of each of these equations represent the population exchange between the upper and lower levels with a reservoir. Taking the difference between the two rate equations, we have
\begin{equation}
\begin{split}
    \frac{d\Delta n}{dt} &= -A_{ul}n_u - 2U_{ul}(t)(B_{ul}n_u-B_{lu}n_l) - (\Gamma_u(t)n_u - \Gamma_l(t) n_l)\\
    &+ (\Lambda_u(t) - \Lambda_l(t)) 
\end{split}
\label{eq:firstrate}
\end{equation}
If $g_u/g_l \sim 1$ and $\Gamma_u(t) \sim \Gamma_l(t)$, eq.\,\ref{eq:firstrate} reduces to
\begin{equation}
    \frac{d\Delta n}{dt} = -A_{ul}n_u - 2U_{ul}(t)B\Delta n + \Gamma(t)\Delta n + \Lambda(t) 
    \label{eq:finrate}
\end{equation}
where $\Lambda(t) = \Lambda_u(t) - \Lambda_l(t)$. The first two terms on the right-hand side represent the radiative coupling between the maser levels, which, in Eq. 7 of \citet{Rajabi2023}, is given by $(i/\hbar)(P^+E^- - E^-P^+)$. For a constant loss rate, the third term can be written as $\mathrm{\Delta n/t_{rel}}$, where $\mathrm{t_{rel}} = \Gamma^{-1}$ is the time scale for the relaxation of the population inversion. The above Eq. \ref{eq:finrate} and Eq. 7 of \citet{Rajabi2023} therefore have the same structure.  

It is obvious that while $\Lambda_u(t)$ and $\Lambda_l(t)$ uniquely determine $\Lambda(t)$ at any time, the inverse problem is highly degenerate. In fact, this is already the case for Eqs.\,\ref{eq:rate1} and \ref{eq:rate2} where a multi-level system has been reduced to a three-level system. In a multi-level system, the individual terms (in the full set of rate equations) that contribute to the loss and pump rates for the upper and lower maser levels explicitly depend on the properties of the radiation field over the entire spectral range. However, information about the radiation field is lost when summing over all the contributing terms to obtain the rate equations for the two masing levels in a three-level system. Reducing the two masing levels further by transforming to $\Delta n = n_u-n_l$ produces, in our opinion, a system which has no connection with the three-level system (Eqs.\,\ref{eq:rate1} and \ref{eq:rate2}) and even less with the original multi-level system, which is subject to a specific time-dependent radiation field (assuming that the flaring is due to changes in radiative pumping). 

The implication of applying the Maxwell-Bloch equations to understand the periodic masers in G962 is that, while an acceptable fit to the flare profiles can be found, it is not possible to uniquely infer the physical pumping process that produced the inversion change. Inspection of the results presented in \citet{Rajabi2023}, sections 5.2 and 5.3, actually illustrates this point. It is noticed that in fitting the flare profiles for the 6.7 GHz masers at 5.0 and 8.0 $\mathrm{km\,s^{-1}}$ as well as the 12.2 GHz maser at 1.7 $\mathrm{km\,s^{-1}}$, the same value of $1.9\times 10^7$s for $\mathrm{t_{rel}}$ (the relaxation time scale for the inversion) was used. The physical implication of using the same value is that the total loss rate out of the $5_1$ (49.1 K above the ground-state) and $6_0$ states for A-methanol is exactly the same as for the $2_0$ (20.1 K above the ground-state) and $3_{-1}$ states of E-methanol for the same pumping radiation field. A- and E-type methanol are two different molecules with different energy level structures, which respond differently to the same radiation field. Even a simple inspection of the radiative coupling of the maser levels (of the 6.7 and 12.2 GHz transitions) to other levels as given, for example, in the Leiden atomic and molecular database \citep{Schoier2005} clearly suggests that achieving the same loss rates for all four states would require a highly specialized radiation field. The fact that the same value of $\mathrm{t_{rel}}$ had to be used to obtain a fit to the decay parts of both the 6.7- and 12.2-GHz flares implies that the decay parts of the flares are not determined by the molecular properties of the two molecules or the external radiation field, but rather by some other external factor.

\subsection{Implications for the colliding-wind binary model}

As with the previous two mechanisms, the CWB model needs to be evaluated within the framework of the above results. To do so, a brief description of the CWB model is provided below. Assuming the presence of a CWB system, strong stellar winds produce hot shocked gas with temperatures on the order 10$^{6-8}$ K, which produces copious amounts of ionizing photons. The intensity of these ionizing photons is modulated due to an eccentric binary system, which results in a pulse of ionizing photons around the periastron passage. These ionizing photons produced from the shocked gas travel outward and cause additional ionization in some volume of partially ionized gas. During the pulse, fractional ionization continues to increase in this volume. At some point after the peak (periastron passage) of the pulse, recombination starts to dominate, at which point "pure" recombination starts and the volume of partially ionized gas returns to a lower state of ionization.  Through this hypothesis, the flaring in the G962 methanol masers is associated with a pulse of ionizing photons produced from the shocked gas around the periastron passage, which increases the ionization in the volume of gas; subsequently, after the pulse, the decay of the flare is due to recombination back to a lower state of ionization. Thus, the maser "sees" the time-dependent change in free-free emission (the source of seed photons, $\propto n_{e}^{2}$) from the partially ionized gas.

Central to the above results is that, in order for the CWB model to be a viable mechanism to explain the periodic methanol masers in G962, it must be able to address the following aspects; the repeatability of the flares, the flare stability, the specific shape of the flare profile, and its duty cycle. A binary system is an ideal astrophysical source for providing a stable and highly repeatable phenomenon. An argument can be made for pulsating stars or periodic accretion, but as mentioned above, it was determined by \citet{Templeton2005} and \citet{Neilson2016} that, in general, the periods of pulsating stars are not constant. Moreover, although periodic accretion \citep{Araya2010} has been proposed to explain the periodic variability of masers, \citet{vanderwalt2016} has already ruled out this possibility. In addition, \citet{vanderwalt2016} also eliminated the spiral shock model of \citet{Parfenov2014}. Nevertheless, the CWB model can describe the repeatability and stability of the flare profile. Although the CWB model can describe the repeatability and stability, it also needs to explain the flare shape and duty cycle of the flare profile.  The flare profile consists of three parts defined in section \ref{sec:results}, (i) the rise time plus the time during flare maximum, (ii) the decay of the flare, and (iii) the quiescent state "between the flares" of the flare profile. 

\begin{figure}[h]
   \centering
   \includegraphics[width=\hsize]{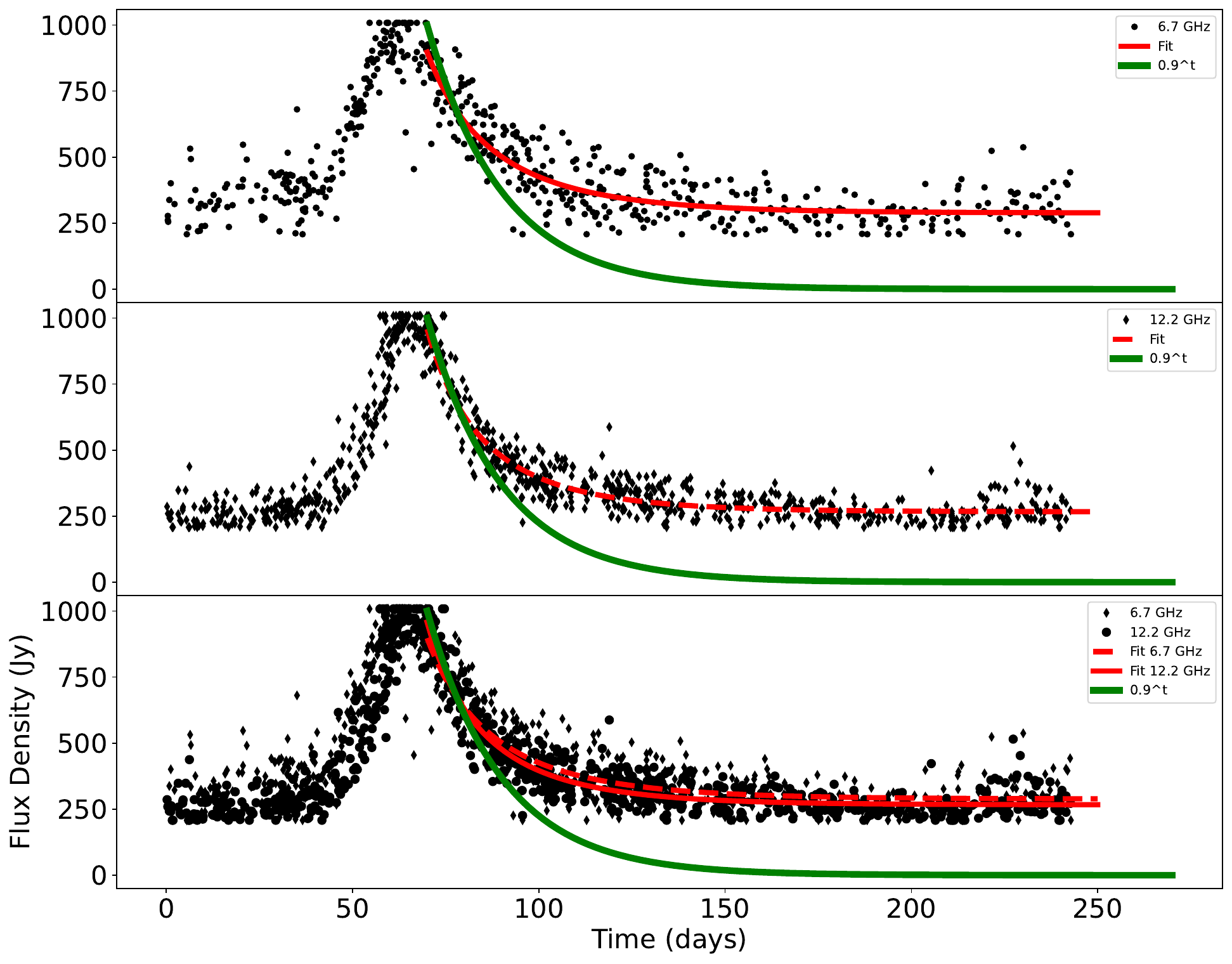}
      \caption{Top panel: The phase profile of the 6.7 GHz methanol masers with the recombination fitted. Middle panel: The phase profile of the 12.2 GHz methanol masers with recombination fitted. Bottom panel: The phase profiles of both 6.7 GHz (cycles) and 12.2 GHz (diamonds), together with recombination fitted. All three panels also show an example fit of an exponential function, which shows that the exponential function only fits well for the first 30 days and not the rest of the decay of the flare. }
         \label{fig:Recomb}
   \end{figure}

Starting with the decay of the flare profile, within the framework of the CWB model, the decay of the flare is described by a partially ionized gas recombining from a higher to a lower state of ionization. In order to test this, the special solution of Eq. A3 from \citet{vanderwalt2009}, given by: 

\begin{equation}
n^{2}_{e}(t) = n^{2}_{e,\star} [\frac{u_{0}+\tanh(\alpha n_{e,\star}t)}{1 + u_{0}\tanh(\alpha n_{e,\star}t)}]^2
\label{eq:recomb}
\end{equation}

where $n_{e,\star}$ is the quiescent electron density, $\alpha$ is the recombination coefficient, and $u_{0}$ = $\frac{n_{e,0}}{n_{e,\star}}$ is the ratio between the quiescent electron density and the maximum electron density ($n_{e, 0}$), which is used to describe the time-dependence of the electron density if only recombination plays a role.  Fig. \ref{fig:Recomb} shows a fit of Eq. \ref{eq:recomb} (red lines) to the data of the 6.7 GHz (top panel), 12.2 GHz (middle panel) and both 6.7 GHz and 12.2 GHz (bottom panel), which fits remarkably well over at least 210 days ($\sim$ 86$\%$ of the period). The derived values for both 6.7 and 12.2 GHz data are $n_{e,\star} \simeq 3.0 \times 10^{5} \mathrm{cm^{-3}}$ and $u_{0}$ $\simeq$ 2.71, thus $n_{e,0} \simeq 8.1 \times 10^{5} \mathrm{cm^{-3}}$. The recombination fits over at least 210 days, suggesting that the flare profile really consists of only 2 intervals, the rise time plus the time during flare maximum (which lasts for $\sim$ 28 days, 11.5$\%$ of the period) and the decay of the flare (which then leaves $\sim$ 215 days, 88.5$\%$ of the flare profile for the decay of the flares) for a period of 243 days.

The green line is a fit of an exponential decay function to the 6.7 GHz and 12.2 GHz methanol maser data, which shows that it only fits for the first 30-40 days of the decay profile. This indicates that the decay of the flare is not described by an exponential decay lifetime but rather that the decay of the flare is fitted remarkably well by pure recombination. In addition, it fits the decay profile for more than 30 flares over a time span of more than 20 years. This suggests the presence of a volume of partially ionized gas capable of increasing its fractional ionization from a lower to a higher state with subsequent recombination back to the lower state. The ionization front (IF) of an \ion{H}{II} region provides such a partially ionized gas capable of changing its time-dependent state of ionization. \citet{sanna2015}, among others, confirm the presence of a \ion{H}{II} region associated with G962, supporting the presence of an IF that can change its time-dependent ionization state. As the partially ionized gas at the IF is most likely optically thin at both 6.7 GHz and 12.2 GHz, it supports the CWB model in which the flare can be explained as the time-dependent change in free-free emission ($\propto n_{e}^{2}$) from the IF of the \ion{H}{II} region. This strongly suggests that at the position of the IF additional ionization is switching on and off, and that the maser is projected against the IF. It supports the CWB model because it provides an additional source of ionizing photons capable of causing the time-dependent change in ionization at the IF. 

The periodic flaring of the methanol masers in G962 can be interpreted as a periastron event in an eccentric CWB system, analogous in principle to the periastron activity observed in WR 140 by \citet{Williams1990}. In WR 140, the highly eccentric orbit causes the stellar separation to decrease drastically near periastron passage, leading to increased wind-collision activity during a small fraction of the orbital period. WR 140 has therefore become a classic example of an eccentric massive binary where the wind-collision region produces phase-dependent observable changes. G962 can be interpreted similarly, the key observational fact is that the methanol masers do not flare randomly; they recur with a stable period. This behaviour is exactly what one would expect if the masers were responding to a periodic periastron event rather than to random changes in the masering gas. The short duty cycle of 0.115 for the flaring of G962 together with the similarity between the flare profiles also suggests an eccentric binary system for G962. The periastron passage can therefore produce a temporary enhancement in ionizing radiation capable of resulting in an enhancement of ionization.  The IF of the \ion{H}{II} region, mentioned above, provides a source for temporary enhancement in ionization and thus free-free emission. An eccentric binary system thus explains the short duty cycle for the observed flaring. In summary, the CWB model explains the decay of the flare profile as a "pure" recombination of the IF of the \ion{H}{II} region from a higher to a lower state of ionization, and the short duration of the flare (which increases the ionization at the IF) can be explained by a pulse of ionizing photons produced around periastron as the result of orbital modulation of the wind-collision region in an eccentric orbit. \citet{vandenheever2019} found that a binary system with an eccentricity of $\epsilon \simeq 0.65-0.70$ is capable of explaining the duty cycle of the flaring and the flare profile. Thus, the CWB model addresses all aspects of the flare profile obtained in this work and explains the periodic flaring of the methanol masers in G962.    

\section{Conclusions}

In the case of G962, and possibly in sources whose flare profiles resemble those of G962, observational evidence strongly indicates that there were no physical changes in the masering region during either the flaring or quiescent phase. This implies that there are only two mechanisms that could be responsible for the periodicity observed, (1) time-dependence of the pumping of the masers and (2) time-dependence of the background seed photons that the masering region amplifies. In the case of (1), \citep{vanderwalt2016} eliminated the proposals of \citep{Parfenov2014} (shocked gas from a spiral shock in a circumbinary disk) and \citep{sanna2015} (a young pulsating massive star). In this article, we have further eliminated that neither (1) FIR pumping effects nor (2) Dicke's superradiance can explain the periodic behavior of the methanol masers. This currently leaves the CWB model as the best possible explanation for the origin and behavior of the periodic masers in G962. Thus, all except one, the CWB model have been eliminated. 

\begin{acknowledgements}

S.P. van den Heever acknowledges the South African Radio Astronomy Observatory (SARAO) for their financial support under the Grant UID: 97954. We also acknowledge and thank Hartebeesthoek Radio Astronomy observatory for providing the data of G009.62+0.20E. The authors also acknowledge the reviewer for their constructive comments that improved the final manuscript.

\end{acknowledgements}

\bibliographystyle{aa} 
\bibliography{ref} 

\end{document}